\input lanlmac.tex
\overfullrule=0pt
\input epsf.tex

\newcount\figno
\figno=0
\def\fig#1#2#3{
\par\begingroup\parindent=0pt\leftskip=1cm\rightskip=1cm\parindent=0pt
\baselineskip=11pt
\global\advance\figno by 1
\midinsert
\epsfxsize=#3
\centerline{\epsfbox{#2}}
\vskip 12pt
{\bf Fig.\ \the\figno:} #1\par
\endinsert\endgroup\par
}
\def\figlabel#1{\xdef#1{\the\figno}}
\def\encadremath#1{\vbox{\hrule\hbox{\vrule\kern8pt\vbox{\kern8pt
\hbox{$\displaystyle #1$}\kern8pt}
\kern8pt\vrule}\hrule}}
\def\omit#1{}

\def\pre#1{{\tt
#1}}

\def\IR{\relax{\rm I\kern-.18em R}}

%
\lref\DIF{P. Di Francesco, 
{\it A refined Razumov-Stroganov conjecture},
JSTAT 2004 P08009,
arXiv:\pre{cond-mat/0407477}.}
\lref\BIBLE{D. Bressoud, {\it Proofs and confirmations. The story of the alternating
sign matrix conjecture}, Cambridge University Press (1999).}
\lref\RS{A.V. Razumov and Yu.G. Stroganov, 
{\it Combinatorial nature
of ground state vector of O(1) loop model},
Theor. Math. Phys. {\bf 138} (2004) 333-337, arXiv:\pre{math.CO/0104216}.}
\lref\RSop{A.V. Razumov and Yu.G. Stroganov, 
{\it O(1) loop model with different boundary conditions and 
symmetry classes of alternating-sign matrices}, 
arXiv:\pre{cond-mat/0108103}.}
\lref\DGR{J. De Gier and V. Rittenberg, {\it Refined
Razumov-Stroganov conjectures for open boundaries}, 
arXiv:\pre{math-ph/0408042}.}
\lref\MNdGB{S. Mitra, B. Nienhuis, J. de Gier and M.T. Batchelor,
{\it Exact expressions for correlations in the ground state 
of the dense $O(1)$ loop model}, arXiv:\pre{cond-math/0401245}}
\lref\Mnosc{S. Mitra and B. Nienhuis, {\it 
Osculating random walks on cylinders}, in
{\it Discrete random walks}, 
DRW'03, C. Banderier and
C. Krattenthaler edrs, Discrete Mathematics and Computer Science
Proceedings AC (2003) 259-264, arXiv:\pre{math-ph/0312036} and
{\it Exact conjectured expressions for correlations in the dense 
O(1) loop model on cylinders}, arXiv:\pre{cond-mat/0407578}.} 
%
%
%
%
\lref\DGR{J. De Gier and V. Rittenberg, {\it Refined
Razumov-Stroganov conjectures for open boundaries}, 
arXiv:\pre{math-ph/0408042}.}
\lref\ROB{D. Robbins, {\it The story of 1,2,7,42,429,7436,...}, 
Mathl. Intelligencer {\bf 13} No.2 (1991) 12-19.}
\lref\OPEN{P. Pearce, V. Rittenberg and J. de Gier,
{\it Critical Q=1 Potts Model and Temperley-Lieb Stochastic Processes},
arXiv:\pre{cond-mat/0108051} and  A.V. Razumov and Yu.G. Stroganov, 
{\it O(1) loop model with different boundary conditions and symmetry 
classes of alternating-sign matrices}, arXiv:\pre{cond-mat/0108103}.}
%
\Title{SPhT-T04/117}
{\vbox{
\centerline{A refined Razumov-Stroganov conjecture II}
}}
\bigskip\bigskip
\centerline{P.~Di~Francesco,} 
\medskip
\centerline{\it  Service de Physique Th\'eorique de Saclay,}
\centerline{\it CEA/DSM/SPhT, URA 2306 du CNRS,}
\centerline{\it F-91191 Gif sur Yvette Cedex, France}
\bigskip
\bigskip\noindent
We extend a previous conjecture \DIF\ relating the Perron-Frobenius eigenvector 
of the monodromy matrix of the O(1) loop model to refined numbers of alternating sign 
matrices. By considering the O(1) loop model on a semi-infinite cylinder
with dislocations, we obtain the generating function for alternating sign matrices
with prescribed positions of $1$'s on their top and bottom rows. This seems to 
indicate a deep correspondence between observables in both models. 


\Date{09/2004}

%
%

\newsec{Introduction: ASM, FPL and the RS conjecture}

Alternating sign matrices (ASM), a classical subject
of combinatorics, have received lots of attention recently in the more
physical context of the fully-packed loop (FPL) model, as well as
in relation to the O(1) loop model, via the celebrated 
Razumov-Stroganov (RS) conjecture.

Alternating sign matrices (ASM) are matrices with entries $0,\pm 1$, such that 
$+1$'s and $-1$'s alternate along each row and column, possibly separated by 
arbitrarily many $0$'s, and such that the entries in each row and column sum to $+1$.
The total number of $n\times n$ ASM reads (see \BIBLE\ for a nice historical exposition
of this formula and many references)
\eqn\asmnum{A_n= \prod_{j=0}^{n-1} {(3j+1)!\over (n+j)!} }
The ASM may be viewed as configurations of the fully-packed loop (FPL) model,
in which each of the edges of a square grid of size $n\times n$ of square lattice may
be occupied by a bond, in such a way that exactly two bonds are incident to 
each vertex, and with the boundary condition that every other external
edge perpendicular to the boundary of the grid is occupied by a bond.
This gives rise to six possible vertex configurations, two of which are ``crossing"
(bonds occupy opposite edges) and four of which are ``turning" (bonds occupy 
consecutive edges). The $0$'s of the ASM correspond to turning vertices of the FPL,
while $\pm 1$'s correspond to crossing vertices.

\fig{Two kinds of observables in the ASM/FPL contexts. An ASM (a) of size $6\times 6$
with a $1$ in position
$5$ in its top row (indicated by an arrow). The corresponding FPL configuration (b), 
with external occupied edges
labeled from $1$ to $12$, has the link pattern 
$(21)(43)(76)(98)(11\, 10)(12\, 5)$.}{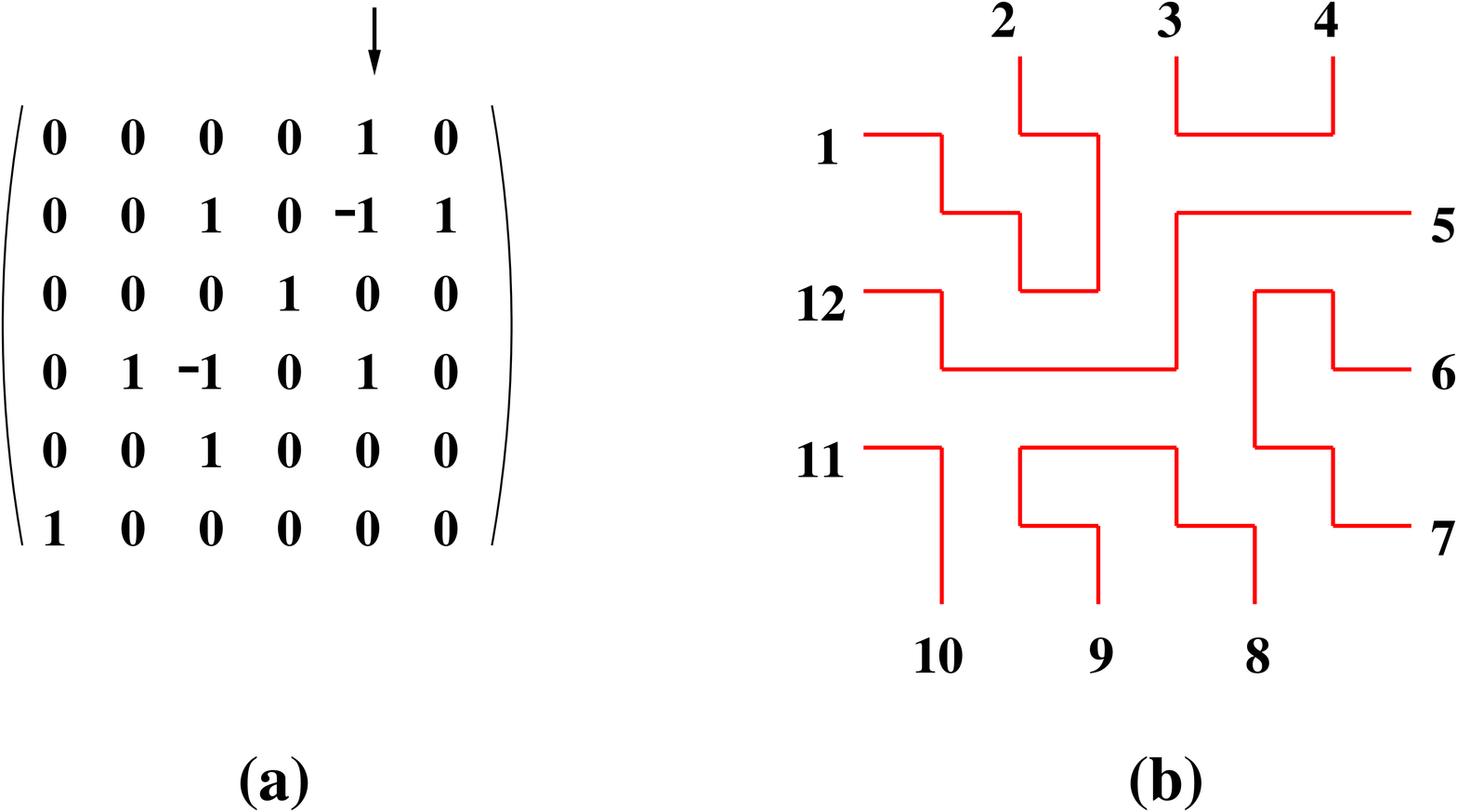}{10.cm}
\figlabel\observables

A natural observable in the context of ASM is the position of the only $1$ present 
in the first row of every such matrix (see Fig.\observables\ (a) for
an illustration). The total number $A_{n,m}$ of ASM of size $n\times n$
with a $1$ in position $m$ (counted from left) in their top row was conjectured
and then proved to read 
\eqn\refasm{A_{n,m}= {n+m-2\choose m-1} {(2n-m-1)!\over (n-m)!} \prod_{j=0}^{n-2} 
{(3j+1)!\over (n+j)!} }

Another natural observable in the context of FPL is the connectivity of external bonds
(see Fig.\observables\ (b) for an illustration).
Indeed, the external bonds are connected by pairs via non-intersecting chains of consecutive bonds
across the grid. After labeling the external bonds
$1,2,...,2n$ clockwise around the grid, these ``planar" pairings are recorded by
link patterns $\pi$, conveniently written as the succession of pairs of connected 
labels clockwise
around the boundary, or equivalently as a chord diagram, connecting the $2n$ points
represented on the boundary of a disk via $n$ non-intersecting
lines or chords. We denote by $LP_n$ the set of link patterns on $2n$ points,
with cardinality $c_n=(2n)!/((n+1)!n!)$, the $n$-th Catalan number.
For each $\pi\in LP_n$, we may ask what is the total number $A_n(\pi)$ of ASM whose FPL 
configurations connect the external bonds according to $\pi$.
Razumov and Stroganov (RS) \RS\ have conjectured a generic answer to this question: the vector
$\Psi_n=(\{A_n(\pi)\}_{\pi\in LP_n})$ is the groundstate eigenvector of the Hamiltonian
of the O(1) loop model, acting on link patterns. 
Many other conjectures have flourished since, involving specific symmetry classes of ASM
and variations on the boundary conditions of the O(1) loop model (see \RSop\ and
also \MNdGB\ \Mnosc\ for a nice review and more conjectures). 

In \DIF, we have been able to mix both types of observables and to find a generalization
of the RS conjecture involving the numbers $A_{n,m}(\pi)$ of ASM with link pattern $\pi$ and
with $1$ in position $m$ in their first row. In the present note, we reformulate this
conjecture in terms of the O(1) loop model on a semi-infinite cylinder, and find that it corresponds
to a dislocation in the lattice. By considering the case of more dislocations, we will 
get other nice conjectures. Another direction of generalization was followed in
\DGR, involving O(1) Hamiltonians with boundary terms, and both approaches look unrelated at
first sight.

This note is organized as follows.
In Sect.2, we review the RS conjecture in the context of O(1) loop model
on a semi-infinite cylinder. Sect.3 is devoted to the reformulation of the refined RS conjecture
of \DIF\ in terms of the O(1) loop model on a semi-infinite cylinder with a dislocation.
In Sect.4, we extend this conjecture to a relation between the two-dislocation O(1) loop model
and the doubly-refined ASM numbers in which the positions of the $1$'s in both top and bottom rows 
of the matrices are recorded. Sect.5 is concerned with the $m$-dislocation case, which produces
new numbers awaiting a good combinatorial interpretation. We gather a few concluding remarks in 
Sect.6.

\newsec{RS conjecture in terms of the O(1) loop gas on a semi-infinite cylinder}

\subsec{The model and its geometrical setting}

We consider a semi-infinite cylinder of (curved) square lattice, with a boundary of 
perimeter $2n$. The configurations of the O(1) loop gas are simply generated
by filling each square face of the lattice by either of the two local
loop configurations $\epsfbox{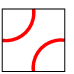}$ or $\epsfbox{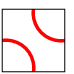}$, say with
weights $t$ and $1-t$ respectively. Indeed, this results
in drawing non-intersecting loops passing by the centers of the edges 
of the lattice, with a fugacity $n=1$ per loop here.

For a given configuration, the centers of the boundary edges, labeled $1,2,...,2n$ 
in clockwise direction,
are pairwise connected by lines of the loop model, according to a 
link pattern $\pi\in LP_n$.

\subsec{Transfer matrix and Hamiltonian}
\fig{Typical configuration of the O(1) loop gas on a semi-infinite cylinder. 
We have isolated the transfer matrix adding an extra row
to the cylinder, and pictured it in terms of face transfer
matrix operators $X_i(t)$, represented as (tilted) squares $\epsfbox{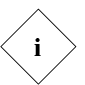}$. 
The last picture involves a gluing along thick red edges, 
corresponding to a trace over an auxiliary space.}{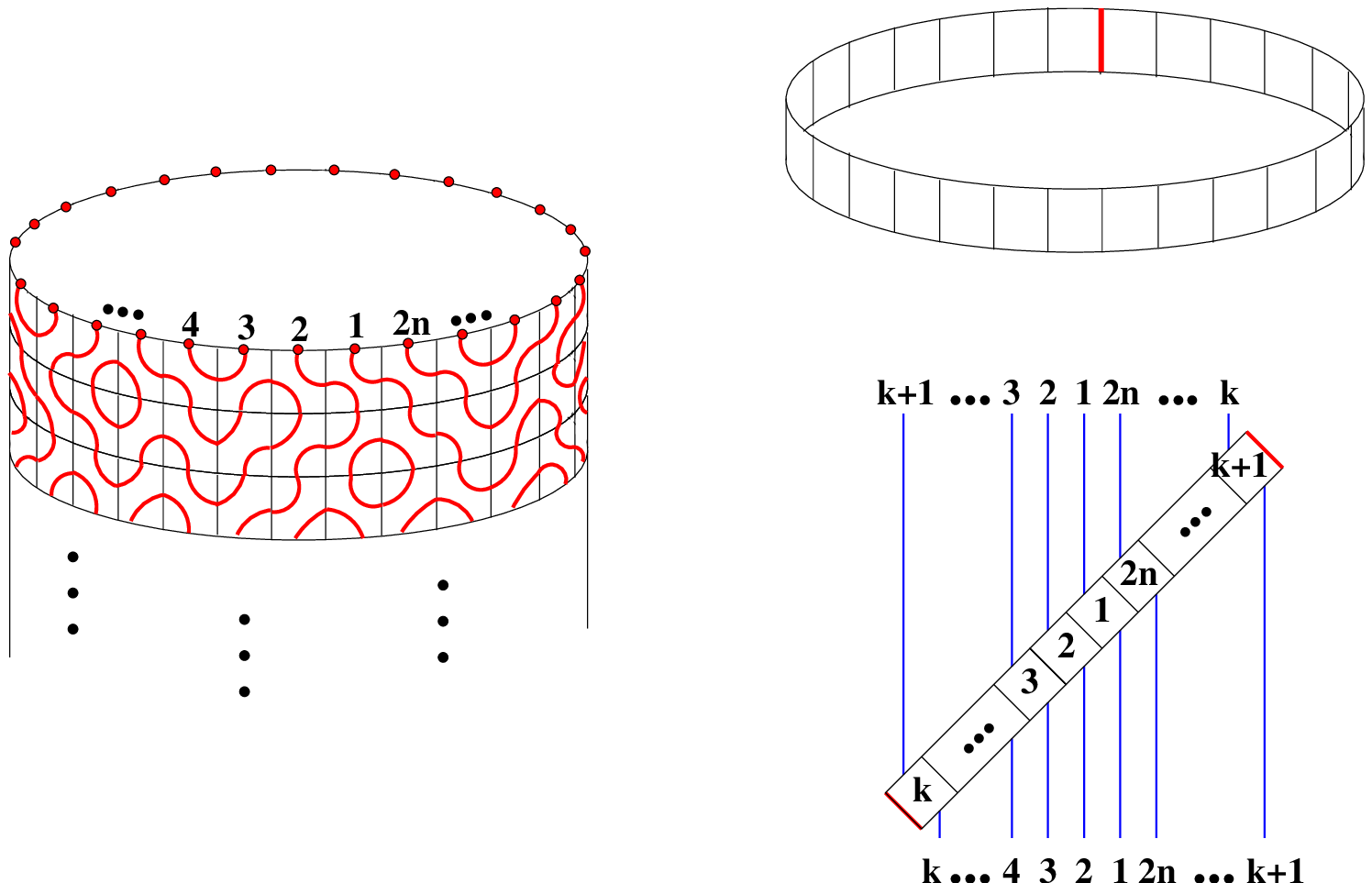}{12.cm}
\figlabel\zeroindent

The configurations of the O(1) loop model are generated by a row-to-row transfer matrix
$T(t)$ including the abovementioned weights, and with periodic boundary conditions
around the cylinder. 
The matrix $T(t)$ has a nice representation in a basis indexed by link patterns,
using the Cyclic Temperley-Lieb algebra generators $e_1,e_2,...,e_{2n}$.
The generator $e_i$ acts on a link pattern $\pi$ as follows: assume the point $i$
is connected to $j$ and $i+1$ to $k$ in $\pi$, then $e_i\pi$ is identical to $\pi$
except that now $j$ is connected to $k$ and $i$ to $i+1$. Note that this is done
without introducing any crossing between chords. The generators satisfy the relations
$e_i^2=e_i$ and $e_ie_{i\pm 1}e_i=e_i$ with the convention that 
$e_{2n+1}\equiv e_1$.
Introducing the face transfer 
matrix operator 
\eqn\ftmo{ X_i(t)=t I + (1-t) e_i}
$T(t)$ is nothing but a product of $X$'s, and the periodic boundary condition
amounts to a trace over an auxiliary space, as displayed in Fig.\zeroindent.

Thanks to the Yang-Baxter relation satisfied by the Boltzmann
weights of the model, itself a consequence of the Temperley-Lieg algebra
relations satisfied by the $e_i$'s, transfer matrices at two distinct values of $t$ commute,
hence they share the same eigenvectors, in particular the same Perron-Frobenius
eigenvector (for $0<t<1$), with maximal eigenvalue and all entries positive.

Interpreting the factor $t$ as a probability, let $\Psi_n(\pi)$
denote the probability in random configurations of the O(1) loop model
(obtained by taking independently on each square face
the weight $\epsfbox{mov1.eps}$ with probability $t$ and $\epsfbox{mov2.eps}$ with probability
$1-t$),
that boundary points of the semi-infinite cylinder be connected according
to a given link pattern $\pi$. These quantities may be viewed as the components of a vector
$\Psi_n=\big(\{\Psi_n(\pi)\}_{\pi\in LP_n}\big)^t$ in the above link pattern basis. 
This vector is easily seen to satisfy
\eqn\pfropsi{ T_n(t) \Psi_n=\Psi_n }
obtained by adding an extra row to the semi-infinite cylinder, and using the action of $T_n(t)$
on link patterns: this simply expresses that adding a row to the semi-infinite cylinder
leaves the probabilities invariant. This shows that $\Psi_n$ is the 
Perron-Frobenius eigenvector of $T_n(t)$, independent of $t$. Expanding $T_n(t)$ around $t=1^-$,
one gets the following null-vector equation:
\eqn\nulvec{H_n  \Psi_n=0 }
where
\eqn\hamiH{H_n=\sum_{i=1}^{2n} (I-e_i)}
is the Hamiltonian of the model.

\subsec{RS conjecture}

The RS conjecture involves a suitably normalized version of $\Psi_n$,
in which all the entries are minimal positive integers, rather than fractions summing 
to $1$ in the case of probabilities. 
By a slight abuse of notation, we still denote this vector
by $\Psi_n$.  
Another characterization is that $\Psi_n(\pi_0)=1$ for the ``parallel" link pattern 
$\pi_0\equiv (2,1)(n+2,n+1)(n+3,n)...(2n,3)$.
Razumov and Stroganov \RS\ conjectured
that the normalized entries $\Psi_n(\pi)$ simply count the number of FPL configurations
on a $n\times n$ grid, paired according to the {\it same} link pattern $\pi$.

A by-product of the RS conjecture is that the sum of entries of $\Psi_n$ should equal the total
number $A_n$ \asmnum\ of ASM of size $n\times n$, namely 
\eqn\asmpsi{v_n \Psi_n=A_n}
where $v_n=(1,1,1...,1)$.

\newsec{First refinement: O(1) loop gas on a semi-infinite cylinder with a dislocation}

\subsec{Geometrical setting}

\fig{Typical configuration of the O(1) loop gas on a semi-infinite cylinder with 
a dislocation, resulting in
an indentation of the boundary. We have isolated the transfer matrix adding an extra row
to the picture, and represented it pictorially in terms of face transfer
matrix operators $X_i(t)$ \ftmo\ as before.}{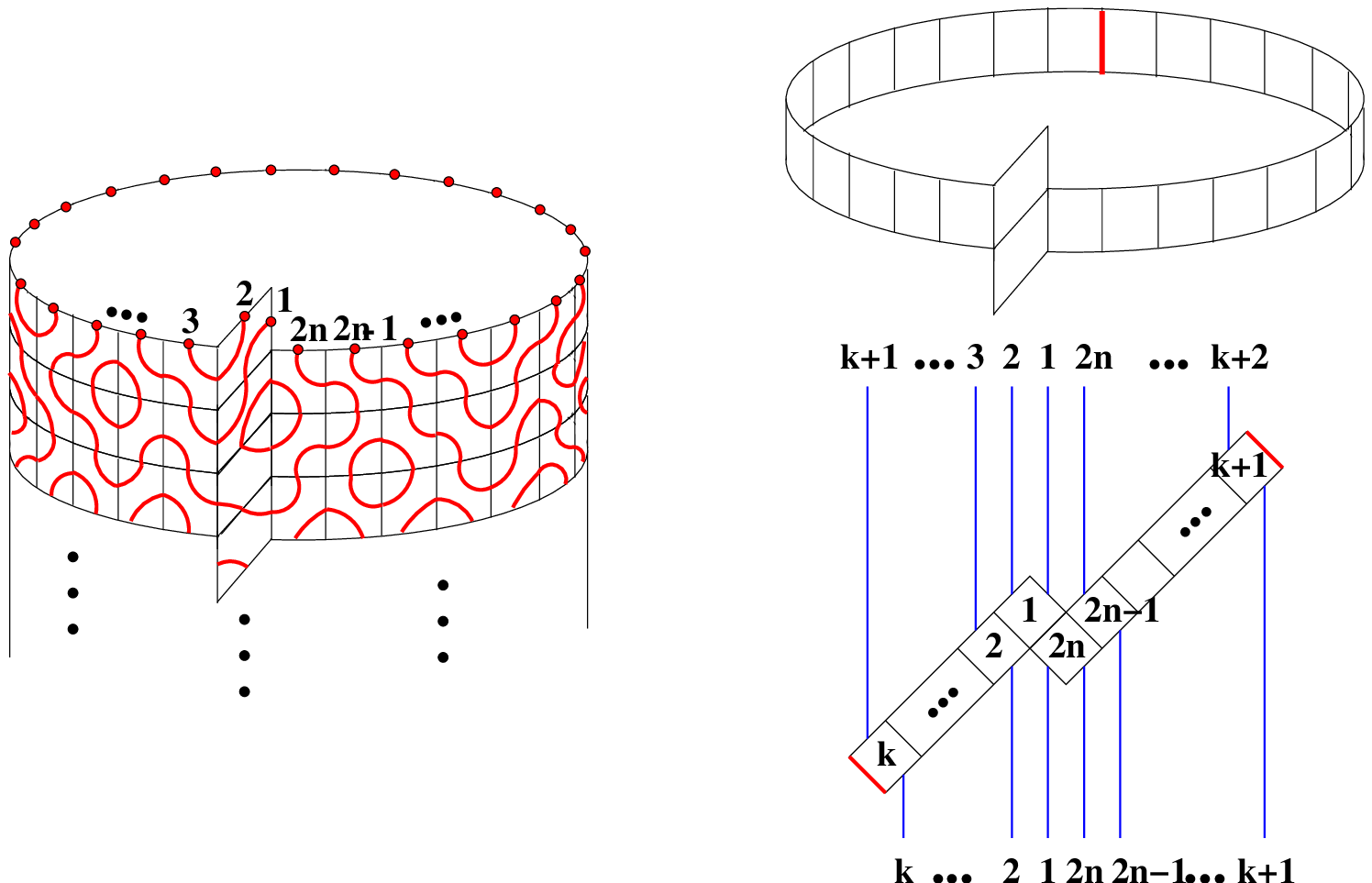}{12.cm}
\figlabel\oneindent

We consider the O(1) loop model on a semi-infinite cylinder of square lattice after
implementing an elementary  dislocation (shift of one lattice spacing)
along an infinite line perpendicular to the boundary, 
as shown in Fig.\oneindent. This is represented as an indentation along the boundary,
whose edge centers are labeled $1,2,...,2n$ in clockwise direction, the label $1$
being attached to the ``dislocated" edge as shown. In any
given O(1) loop model configuration, these $2n$ points are paired according to 
a link pattern $\pi$.

Like in Sect.2.2, we consider the probability $\Psi_n^{(1)}(\pi \vert t)$ that a random 
configuration of the loop model connects the boundary points according
to the link pattern $\pi\in LP_n$.

\subsec{Transfer matrix and refined RS conjecture}

The configurations of the model may be generated by use of a transfer matrix 
$T_n^{(1)}(t)$, which corresponds to adding a ``row" to the picture, as shown in 
Fig.\oneindent.  This matrix is expressed simply as
\eqn\simpext{ T_n^{(1)}(t)=X_1(t)X_2(t)...X_{2n}(t) }
as should be clear from Fig.\oneindent\ (the trace over the auxiliary space, 
indicated by the thick red line, is now an ordinary contraction in the product of the $X$'s).
Forming again a vector $\Psi_n^{(1)}(t)=\big(\{\Psi_n^{(1)}(\pi \vert t)\}_{\pi \in  LP_n}\big)^t$, we obtain
the Perron-Frobenius eigenvector equation
\eqn\pfoneind{ T_n^{(1)}(t) \Psi_n^{(1)}(t)=\Psi_n^{(1)}(t) }
expressing that adding a row to the indented semi-infinite cylinder 
does not affect the probabilities.
Note that $v_n$ is the left Perron-Frobenius eigenvector of $T_n^{(1)}(t)$ 
with same eigenvalue $1$,
due to $v_ne_i=v_n$ (as $e_i$ sends any link pattern to another one) 
and hence $v_n X_i(t)=X_i(t)$ for all $i$.
In \DIF, picking a normalization in which $\Psi_n^{(1)}(\pi_0\vert t)=1$, it  was conjectured
that all entries $\Psi_n^{(1)}(\pi|t)$ are polynomials of $t$ with degree $\leq n-1$,
and that
\eqn\conjdif{v_n\Psi_n^{(1)}(t)=\sum_{m=1}^n A_{n,m}t^{m-1}}
the generating function for the numbers $A_{n,m}$ ASM with a $1$ at a fixed position $m$
in their top row \refasm.

Introducing the clockwise rotation of link patterns 
\eqn\rotalp{\eqalign{ r:&\ \ i \to i+1 \qquad {\rm for}\  i=1,2,...,2n-1\cr 
r:&\ \ 2n\to 1\cr}}
it was also found that for each link pattern $\pi$,
the sums $\sum_{m=0}^{\ell(\pi)-1} \Psi_n(r^m \pi|t)$, $\ell(\pi)$ the smallest positive integer
$\ell$ such that $r^\ell \pi=\pi$, are nothing but the generating functions of FPL configurations
with a fixed position of the crossing vertex in their top row, and with connectivities given by
$\pi$ up to rotations.

\subsec{A family of commuting operators}

The rotation $r$ \rotalp\ of link patterns introduced above satisfies the relations
$r e_i =e_{i-1} r$ for all $i$, and therefore $r X_i(t) =X_{i-1}(t)r$.
Introducing the operators
\eqn\defui{ U_i(t)= r^{i} X_{2n+1-i}(t)X_{2n+2-i}(t)\cdots X_{2n}(t) }
we immediately get that
\eqn\comu{ U_i(t)U_j(t)=U_j(t)U_i(t)=U_{i+j}(t)}
by commuting the $r$'s all the way to the left. 
So all the $U$'s commute with each other and in particular, $U_{2n}(t)=T_n^{(1)}(t)$ commutes
with $U_1(t)=r X_{2n}(t)$, therefore these two operators share the same 
Perron-Frobenius eigenvector $\Psi_n^{(1)}(t)$. 
This provides us with an alternative characterization of $\Psi_n^{(1)}(t)$ via
\eqn\carpsi{U_1(t)\Psi_n^{(1)}(t)=\Psi_n^{(1)}(t) \ \ \Leftrightarrow \ \ H_n^{(1)}(t)\Psi_n^{(1)}(t)=0}
where
\eqn\defH{H_n^{(1)}(t)=I-r^{-1}+(t-1)(I-e_{2n})}
The latter looks much simpler than $T_n^{(1)}(t)-I$, in particular it is linear in $t$.

\fig{The action with $U_i(t)$ of \defui\ on configurations of the O(1) loop model
on a semi-infinite cylinder. Top: the product of $X$'s adds up $i$ squares to the boundary, thus shifting
the indentation by $i$ lattice spacings (the edge of the new dislocation is marked by a red dot).
Bottom: the operator $r^{i}$ shifts the indentation back to the initial position
(red dot). The boundary and therefore the probabilities of the model are left invariant
in the process.}{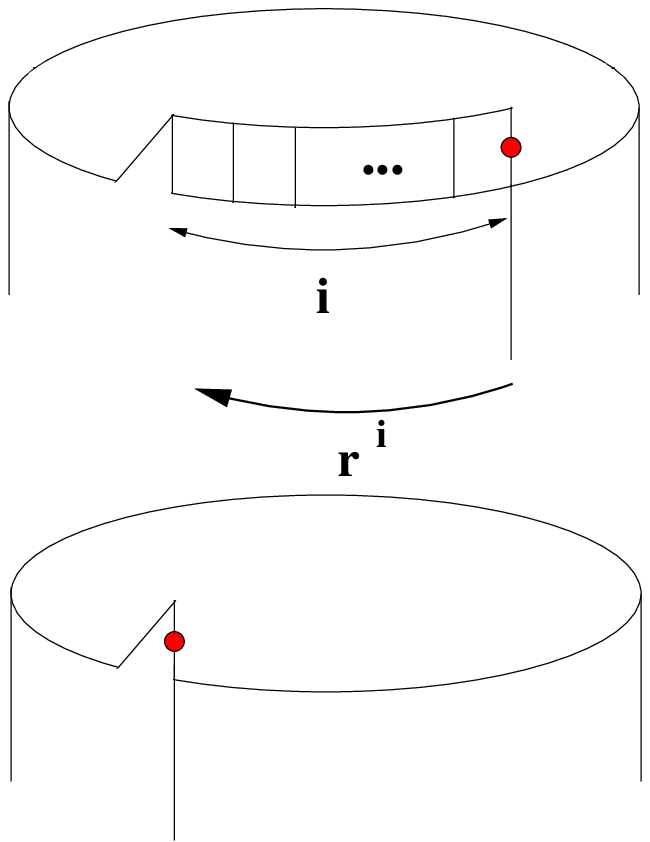}{5.cm}
\figlabel\Ui

The fact that $U_i(t)\Psi_n^{(1)}(t)=\Psi_n^{(1)}(t)$ for all $i$ may be understood directly from
a geometrical point of view, as shown in Fig.\Ui. The multiplication by $X_{2n+i-i}(t)...X_{2n}(t)$ 
amounts to adding a row of $i$ consecutive
squares to the semi-infinite cylinder, from some position down to the indentation. This has the net
effect of shifting the indentation of the boundary by $i$ lattice spacings. Finally, the operator $r^{i}$
rotates back the picture in such a way that the indentation gets back to the initial position: the 
probabilities, and therefore $\Psi_n^{(1)}(t)$, are clearly invariant in the process.

\newsec{Second refinement: O(1) loop gas on a semi-infinite cylinder with two dislocations}

\subsec{Geometrical setting}

\fig{A typical ``two-dislocation" boundary condition on the semi-infinite
cylinder. The transfer matrix is shown and then decomposed into a product of $X_i(t)$
operators as before.}{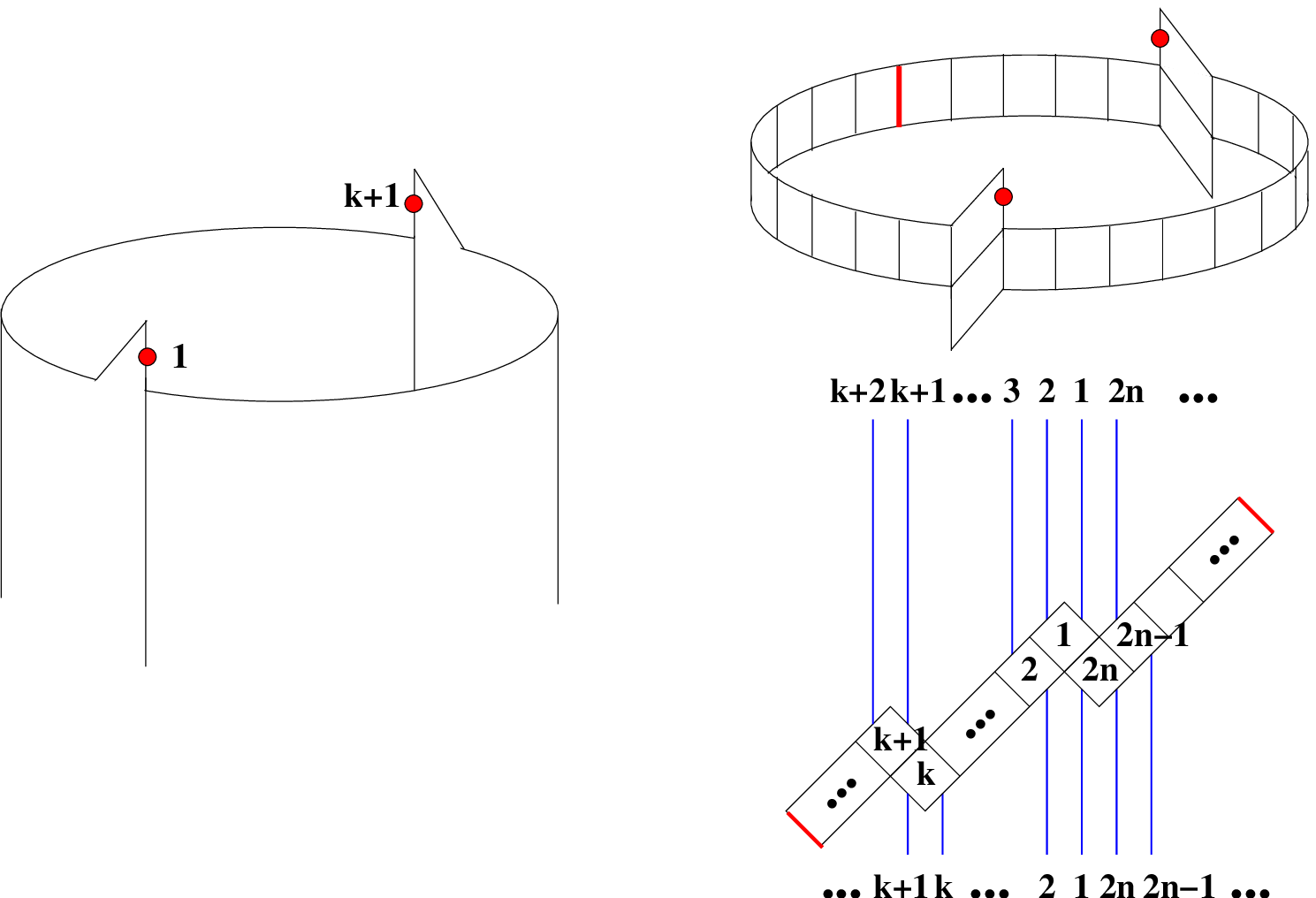}{12.cm}
\figlabel\twoindent

Inspired by the interpretation of Sect.3, we now consider the O(1) loop model
on a semi-infinite cylinder
of square lattice with $2n$ boundary edges, and with {\it two} elementary dislocations, 
as shown in Fig.\twoindent, say with
dislocated edges labeled $1$ and $k+1$. We denote by $\Psi_n^{(1,k+1)}(\pi\vert t)$ the
probability that a random O(1) loop model configuration on this semi-infinite cylinder
connects the boundary points according to the link pattern $\pi\in LP_n$.

\subsec{Transfer matrix, Perron-Frobenius eigenvector and a first conjecture}

It is easy to read off Fig.\twoindent\ the following transfer matrix $T_n^{(1,k+1)}(t)$
that adds up a row to the semi-infinite cylinder with two dislocations:
\eqn\twoimat{ T_n^{(1,k+1)}(t)= X_{k+1}(t)X_1(t)X_2(t)...X_k(t)X_{k+2}(t)X_{k+3}(t)...X_{2n}(t)}
and the vector $\Psi_n^{(1,k+1)}(t)=\{ \Psi_n^{(1,k+1)}(\pi\vert t)\}_{\pi \in LP_n}$
is nothing but the Perron-Frobenius eigenvector of $T_n^{(1,k+1)}(t)$:
\eqn\pftwo{ T_n^{(1,k+1)}(t) \Psi_n^{(1,k+1)}(t)=\Psi_n^{(1,k+1)}(t) }

Like in Sect.3.3, we may drastically simplify the eigenvector equation \pftwo\ by noticing
that we may replace $T_n^{\{1,k\}}(t)$ with any of the operators 
\eqn\utwo{ U^{(1,k+1)}_i(t)= r^{i} X_{2n+1-i}(t)X_{k+1-i}(t)X_{2n+2-i}(t)
X_{k+2-i}(t)...X_{2n}(t)X_{k}(t)} 
which simply add up successively $i$ consecutive pairs of squares, one after
each of the two indentations, 
thus shifting both of them by $i$ lattice spacings, and then takes them both to 
their original positions by the global rotation $r^{i}$. 
As in Sect.3.3, we find that all these operators commute with each other as 
$U^{(1,k+1)}_i(t)U^{(1,k+1)}_j(t)=U^{(1,k+1)}_{i+j}(t)$, and they all leave
the probabilities invariant. Note that none of the $U^{\{1,k\}}_i(t)$ is actually 
equal to the transfer matrix $T_n^{(1,k+1)}(t)$, as opposed to the case of Sect.3.3, 
although they do commute with it.
The simplest of all equations obeyed by $\Psi_n^{(1,k+1)}(t)$ reads
\eqn\simplu{ U^{(1,k+1)}_1(t)\Psi_n^{(1,k+1)}(t)=\Psi_n^{(1,k+1)}(t) \ \ \Leftrightarrow \ \ 
H_n^{(1,k+1)}(t) \Psi_n^{(1,k+1)}(t)=0 }
where
\eqn\defHtwo{H_n^{(1,k+1)}(t)=X_k(t)X_{2n}(t)-r^{-1}
=I-r^{-1}+(1-t)(e_k+e_{2n}-2I)+(1-t)^2(e_k-I)(e_{2n}-I)}
It is simply quadratic in $t$.

A direct computation of the first few $\Psi_n^{(1,k+1)}(t)$ has led us to the following conjecture.
With a suitable normalization described below, the sum of the components of $\Psi_n^{(1,k+1)}(t)$
is a polynomial of degree $2(n-1)$ with only non-negative integer coefficients, independent of
$k\in \{ 2,3,...,2n-2\}$. This polynomial coincides with the generating function for the numbers
of ASM with fixed sum of the positions of their $1$'s in the top and bottom row.
More precisely, let $A_{n,m,p}$ denote the total number
of ASM with a $1$ in position $m$ (counted from the left) in the top row and a $1$ in position $p$
(counted from the right) in the bottom 
row\foot{These numbers are displayed in appendix A up to $n=6$, via the generating polynomials
$A_n(t,u)=\sum_{m,p=1}^{n} A_{n,m,p}t^{m-1}u^{p-1}$.
As opposed to the refined ASM numbers \refasm, 
these doubly-refined numbers do not seem to have any nice 
product form.
As an example, the ASM of Fig.\observables\ (a) contributes by $1$
to $A_{6,5,6}=105$ (coefficient of $t^4 u^5$ in $A_6(t,u)$ of eq.(A.1)), 
as the $1$ in its bottom row sits in position $6$ counted from the right.}, then we have
\eqn\sumofentwo{ v_n \Psi_n^{(1,k+1)}(t)=\sum_{m,p=1}^{n} A_{n,m,p} t^{m+p-2} }
That this quantity is independent of $k$ is a direct consequence of the fact that $v_n$ is
a left eigenvector of all the $X$'s with eigenvalue $1$, namely $v_nX_i(t)=v_n$ for all $i$.
Indeed, for $l<k$, we have $\Psi_n^{(1,l+1)}(t)=X_{l+1}(t)X_{l+2}(t)...X_k(t)\Psi_n^{(1,k+1)}(t)$
for the vectors of probabilities,
where the indentation in position $k+1$ has been shifted by $k-l$ lattice spacings counterclockwise,
and therefore $v_n \Psi_n^{(1,l+1)}(t)=v_n\Psi_n^{(1,k+1)}(t)$.
The normalization of all the $\Psi$'s is then fixed by that of $\Psi_n^{(1,n+1)}(t)$, in which we fix
$\Psi_n^{(1,n+1)}(\pi_0\vert t)=1$. With this choice, it turns out that all the components
of $\Psi_n^{(1,n+1)}(t)$ are polynomials of $t$ of degree $\leq 2n-2$, with only non-negative integer
coefficients, and summing to \sumofentwo.

\subsec{Doubly refined RS conjecture}

\fig{The double-helix decomposition of the two-dislocation semi-infinite cylinder
of square lattice (glued here along the thick red line as indicated). 
Blue strips correspond to drawing the two possible face loop configurations
with probabilities $t$ and $1-t$,
while pink strips correspond to probabilities $u$ and $1-u$.}{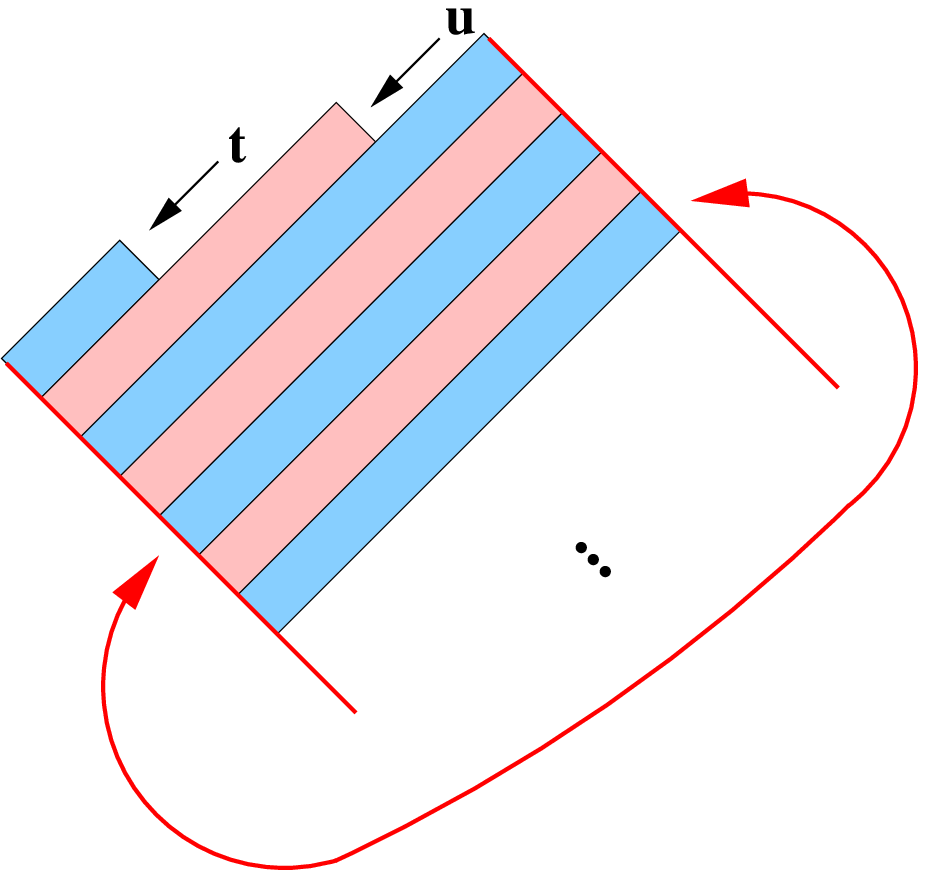}{8.cm}
\figlabel\doubhelix

The result of the previous section suggests to look for a further refinement 
of the RS conjecture, allowing for
recovering the doubly-refined ASM numbers $A_{n,m,p}$ individually in the context 
of the O(1) loop model. This is
done as follows. 
Keeping the geometry of the cylinder as in Sect.4.1, we consider an inhomogeneous
random O(1) loop model in which the loop configurations 
$\epsfbox{mov1.eps}$ and $\epsfbox{mov2.eps}$, are drawn
with probabilities $t$ and $1-t$ respectively on certain faces of the lattice, and $u$ and $1-u$
on others. We actually decompose the two-dislocation semi-infinite cylinder of square lattice,
according to Fig.\doubhelix, into a double-helix formed with one strip of each kind of face.

As before, assuming that the two indentations are in positions $1$ and $k+1$, we denote by 
$\Psi_n^{(1,k+1)}(\pi \vert t,u)$ the probability that a random
configuration of this inhomogeneous O(1) model connects the boundary points according to the
link pattern $\pi$, and we form the vector 
$\Psi_n^{(1,k+1)}(t,u)=\big(\{\Psi_n^{(1,k+1)}(\pi \vert t,u)\}_{\pi\in LP_n}\big)^t$.
The transfer matrix
for the model must involve the addition of $4n$ squares (and make two complete turns), however
the vector $\Psi_n^{(1,k+1)}(t,u)$ is invariant under the action of the operators 
\eqn\uoftu{U^{(1,k+1)}_i(t,u)=r^i X_{k+1-i}(u)X_{2n+1-i}(t)X_{k+2-i}(u)X_{2n+2-i}(t)...
X_{k}(u)X_{2n}(t)}
for all $i$, so we may write the simplest of the corresponding Perron-Frobenius equations
$U^{(1,k+1)}_1(t,u)\Psi_n^{(1,k+1)}(t,u)=\Psi_n^{(1,k+1)}(t,u)$, or equivalently
\eqn\pfrosimp{ H^{(1,k+1)}_n(t,u) \Psi_n^{(1,k+1)}(t,u)=0}
with 
\eqn\defhtwotu{ H^{(1,k+1)}_n(t,u)=I-r^{-1}+(1-t)(e_{2n}-I)+(1-u)(e_k-I)+(1-t)(1-u)(e_k-I)(e_{2n}-I)}
This operator is simply linear in each of the variables $t,u$. Note that when $u=1$
\defhtwotu\ reduces to \defHtwo.
We have computed the first few $\Psi_n^{(1,k+1)}(t,u)$ by use of eqs.\pfrosimp-\defhtwotu.
The results have suggested the following conjecture.

Picking a normalization in which $\Psi_n^{(1,n+1)}(\pi_0\vert t,u)=1$, the entries
of $\Psi_n^{(1,n+1)}(t,u)$ are all polynomials of $t,u$ of degree $\leq n-1$ in each variable,
with only non-negative integer coefficients.
Moreover the sum of entries is nothing but the generating function for ASM with $1$'s in fixed
positions on their top and bottom row:
\eqn\sumtu{ v_n \Psi_n^{(1,n+1)}(t,u) =\sum_{m,p=0}^{n-1} A_{n,m,p} t^{m-1} u^{p-1} }
The vectors $\Psi_n^{(1,n+1)}(t,u)$ are listed in appendix B up to $n=5$, while the
generating polynomials for ASM with $1$'s in fixed
positions on their top and bottom row are listed in appendix A up to $n=6$.

We have not been able to relate directly the components of $\Psi_n^{(1,n+1)}(t,u)$ to 
FPL configurations with fixed connectivity and positions of crossing vertices in their
top and bottom rows.

Noticing again that we may sweep all the vectors $\Psi_n^{(1,k+1)}(t,u)$ upon acting with $X$'s
on any of them, we find that eq.\sumtu\ holds for any $\Psi_n^{(1,k+1)}(t,u)$, for $k=2,3,...,2n-2$,
although entries are no longer necessarily polynomials with non-negative integer coefficients.
The case $k=n$ is singled-out as corresponding to the only ``self-reflected" boundary, namely allowing
for a gluing between the original semi-infinite cylinder and its reflection, into an
infinite cylinder expressed as a glued infinite double-helix of the two types of faces.

\newsec{Multiple refinements: O(1) loop gas on a semi-infinite cylinder with
arbitrarily many dislocations}

The results of Sect.4 lead to the following straightforward generalization. We now consider
a semi-infinite cylinder with $m$ dislocations ($m\leq n$), with $m$ indented 
boundary edges labeled $1,k_1+1,k_1+k_2+1,...,k_1+k_2+...+k_{m-1}+1$ clockwise.  
Taking uniform probabilities $t$ and $1-t$ for the
two possible O(1) loop face configurations, 
we denote by 
$\Psi_n^{(\{k\})}(\pi\vert t)\equiv 
\Psi_n^{(1,k_1+k_2+...+k_{m-1}+1,k_1+k_2+...+k_{m-2}+1,...,k_1+1)}(\pi\vert t)$
the probability that a random configuration of the O(1) loop model connects the boundary
points according to the link pattern $\pi\in LP_n$.

As before, we readily get the simplest Perron-Frobenius equation for the
vector $\Psi_n^{(\{k\})}(t)=\big(\{ \Psi_n^{(\{k\})}(\pi\vert t)\}_{\pi\in LP_n}\big)^t$
\eqn\pfrosimplissimus{ U^{(\{k\})}_1(t)\Psi_n^{(\{k\})}(t)=\Psi_n^{(\{k\})}(t)}
by acting with the operator
\eqn\redily{ U^{(\{k\})}_1(t)=r X_{k_1+k_2+...+k_{m-1}}(t) X_{k_1+k_2+...+k_{m-2}}(t)...X_{k_1}(t)X_{2n}(t)}
which shifts all $m$ indentations by one lattice spacing and then takes them back to their original
position via a global clockwise rotation $r$.
The operator \redily\ has degree $m$ in $t$. Using eqs.\pfrosimplissimus-\redily, we have computed
the first few vectors $\Psi$, and the results have suggested the following conjecture.

As before, with a suitable normalization of the $\Psi$'s, the sum of entries
$P_n(m\vert t)\equiv v_n\Psi_n^{(\{k\})}(t)$ is independent of the $k_i$'s, 
provided they are all $\geq 2$, and it is
a polynomial of degree $nm-[{m^2+1\over 2}]$ of $t$, with only non-negative integer coefficients.
Moreover, some specific choices of the $k$'s (those which best spread the positions of the indentations
around the boundary, and correspond like in Sect.4.3 to ``self-reflected" half-cylinder boundaries) 
yield $\Psi$'s whose entries are all polynomials
of $t$ with only non-negative integer coefficients, but we have not been able to relate directly the
components of these $\Psi$'s to specific classes of FPL configurations.
In particular, when the number of dislocations is maximal ($m=n$) and all $k_i=2$, 
we find that all the entries
of $\Psi$ are polynomials of degree $\leq [{n^2\over 2}]$ with non-negative integer 
coefficients.
The first few $P_n(m\vert t)$'s for $1\leq m\leq n\leq 6$ are listed in appendix C below.

By construction, we have $P_n(m\vert t=1)=A_n$ the total number of FPL or ASM of size $n\times n$.
We may therefore expect that the polynomials $P_n(m\vert t)$ yield decompositions of these numbers according
to some properties of the corresponding ASM or FPL configurations. In Sects.3.2 and 4.2, we have found
such interpretations for $P_n(1\vert t)$ and $P_n(2\vert t)$ respectively.

Our attempts to decorate the $m$-dislocation case with different probabilities corresponding
to multiple helix decompositions of the semi-infinite cylinder have not led
us to any nice multi-variable polynomial generalizations of $P_n(m\vert t)$: 
it seems that the two-variable generalizations \sumtu\ are the best we can do.

\newsec{Conclusion and discussion}

\subsec{RS conjectures and the ASM/FPL--TSSCPP correspondence}

In this note we have extended the refined RS conjecture of \DIF\ by rephrasing it in terms of the
O(1) loop model on a semi-infinite cylinder. This has led to other generalizations in which we 
consider a cylinder with defects (dislocations), translating into indentations on the boundary.
The classical RS conjecture corresponds to the case of no dislocation. The refined RS conjecture of
\DIF\ corresponds to the case of one dislocation. 

By decorating the two-dislocation case with two types of probability weights corresponding to a
decomposition of the cylinder into a double-helix, we have found that the sum of entries
of the suitably renormalized probability vector of the model could be identified with
the generating function for doubly-refined ASM numbers, namely the numbers of ASM with 
both positions of their $1$'s in the top and bottom row specified.
These numbers were already referred to in \ROB\ in the context of the (still mysterious) relation
between ASM and totally-symmetric self-complementary plane partitions (TSSCPP). 
More precisely, the generating functions for TSSCPP with both numbers $f_2$ and $f_3$ fixed
(c.f. \ROB\ for definitions) also coincide with our polynomials $v_n \Psi_n^{(1,n+1)}(t,u)$.
Our two-dislocation conjecture may be a first step toward a more precise
connection between all these counting problems. Actually, it might be that the apparent mismatch
between the entries $\Psi_n^{(1,n+1)}(\pi\vert t,u)$ and the doubly-refined ASM number generating
functions with fixed connectivity $\pi$, $\sum_{m,p} A_{n,m,p}(\pi) t^{m-1}u^{p-1}$, even
up to rotations of $\pi$, simply means that we have not yet understood the correct ASM 
picture corresponding to $\Psi_n^{(1,n+1)}(t,u)$. In particular, in \ROB, Robbins
describes three numbers $f_1,f_2,f_3$ attached to each TSSCPP, any pair of which have the same
distribution within TSSCPP as the $A_{n,m,p}$ within ASM. If we believe in the existence of 
a natural ASM-TSSCPP bijection, we see that a third observable is clearly missing in the ASM
picture to account for the triplet $f_1,f_2,f_3$. This third observable is perhaps the
missing link between our vector $\Psi_n^{(1,n+1)}(t,u)$ and (triply?) refined ASM with fixed
connectivities.

In addition, we have been able to produce more non-negative integer numbers out of the 
$m$-dislocation case, which, we hope, should have an interpretation in both ASM and TSSCPP contexts. 

\subsec{Open case}
\fig{O(1) model on a semi-infinite strip of width $2n$ (open cylinder), 
with $p$ indentations on the boundary (red dots).
The corresponding transfer matrix is decomposed into strips of tilted squares, corresponding
to multiplication by operators $X_i(t)$.}{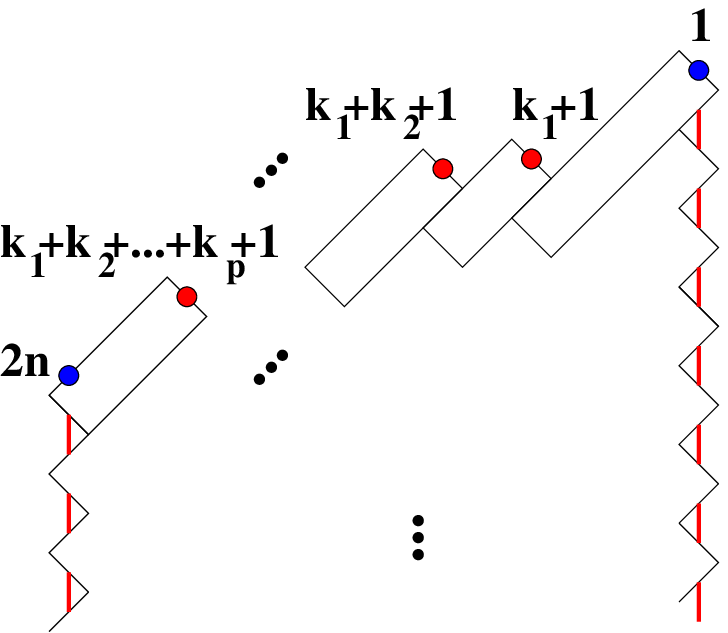}{8.cm}
\figlabel\open

The conjectures of this paper have a natural extension in the open (or semi-infinite strip) case, 
where we draw an impenetrable half-line along the cylinder that separates the points $1$ and $2n$ 
of the boundary, thus transforming the domain into a semi-infinite strip of width $2n$.  
Going directly to the most general situation of a boundary with $p$ indentations
in positions $\{k\}\equiv k_1+1,k_1+k_2+1,\cdots,k_1+k_2+\cdots+k_p+1$
(see Fig.\open), and introducing the probabilities $\Psi_{2n}^{O\,\{k\}}(\pi\vert t)$ that 
a random homogeneous O(1) loop model configuration connects the boundary points according to 
the link pattern $\pi\in LP_n$, we get the eigenvector equation
\eqn\eigopen{ T_{2n}^{O\,\{k\}}(t) \Psi_{2n}^{O\,\{k\}}(t)=\Psi_{2n}^{O\,\{k\}}(t)}
expressing the invariance of the probabilities under addition of a layer of squares
parallely to the boundary, via the transfer matrix
\eqn\tmatopen{\eqalign{&T_{2n}^{O\,\{k\}}(t)=
X_{k_1+...+k_p+1}(t) X_{k_1+...+k_{p-1}+1}(t)...X_{k_1+1}(t) 
\times X_1(t)X_2(t)...X_{k_1}(t)\cr
&\times X_{k_1+2}(t) X_{k_1+3}(t) ... X_{k_1+k_2}(t)\times ...\times 
X_{k_1+...+k_p+2}(t)X_{k_1+...+k_p+3}(t)...X_{2n}(t)\cr}}
The computation of the first few $\Psi^O$'s has led us to the following conjecture.

Noting that all the eigenvectors $\Psi_{2n}^{O\,\{k\}}(t)$ at different values and numbers of $k$'s are 
all related to one-another via products of $X$'s, 
we find that in a suitable global normalization of all these vectors, the sum of entries
\eqn\sumeven{ P_{2n}^O(t)= v_n \Psi_{2n}^{O\,\{k\}}(t) }
is a polynomial of $t$ independent of $p$ and the $k$'s, of degree $n(n-1)$
and with only non-negative integer coefficients. 

This is readily extended to the case of a strip of width $2n-1$, by simply removing 
the factor $X_{2n}(t)$ in the definition of $T_{2n}^{O\,\{k\}}(t)\to T_{2n-1}^{O\,\{k\}}(t)$.
The conjecture above becomes that in a suitable normalization, the sum of eigenvector entries
\eqn\sumodd{ P_{2n-1}^O(t)=v_n \Psi_{2n-1}^{O\,\{k\}}(t)}
is a polynomial of $t$ independent of
$p$ and the $k$'s, of degree $(n-1)^2$ and with only non-negative integer coefficients.

The polynomials $P_m^O(t)$ are listed in appendix D, up to $m=10$. Note that 
according to the classical RS conjectures for the open case \OPEN, 
$P_{2n}^O(t=1)$ is the total number $A_V(2n+1)$ of vertically-symmetric ASM/FPL of size 
$(2n+1)\times (2n+1)$, while $P_{2n-1}^O(t=1)$ is the total number $N_8(2n)$
of cyclically-symmetric transpose-complement plane partitions in a box of size 
$2n\times 2n\times 2n$.  The polynomials $P_m^O(t)$ form respective refinements 
of these numbers, which still await some good combinatorial interpretation.

\bigskip

\noindent{\bf Acknowledgments}

We thank E. Guitter, C. Krattenthaler and J.-B. Zuber for their help at different 
stages of this work.
\vfill\eject
\appendix{A}{Polynomials generating the numbers of ASM with fixed positions of $1$'s
in their top and bottom row}
We list below up to $n=6$ the polynomials $A_n(t,u)=\sum_{m,p=1}^n A_{n,m,p}t^{m-1} u^{p-1}$,
which generate the numbers $A_{n,m,p}$ of ASM with $1$'s in their top and bottom rows
respectively sitting at positions $m$ and $p$ counted from left (top) and right (bottom).
These were computed by transfer matrix techniques on the FPL model.
\eqn\polrefref{\eqalign{
A_1(t,u)&=1\cr
A_2(t,u)&=1\cr
&+t u\cr
A_3(t,u)&=1+t\cr
&+u+tu+t^2u\cr
&+tu^2+t^2u^2\cr
A_4(t,u)&=2 + 3 t + 2 t^2 \cr
&+ 3 u + 5 t u + 4 t^2 u + 2 t^3 u \cr
&+ 2 u^2 + 4 t u^2+ 5 t^2u^2 + 3 t^3 u^2 \cr
&+ 2 t u^3 +3 t^2 u^3 + 2 t^3 u^3\cr
A_5(t,u)&=7 + 14 t + 14 t^2 + 7 t^3 \cr
&+ 14u + 30 t u + 33 t^2 u + 21 t^3 u + 7 t^4 u \cr
&+ 14 u^2+ 33 t u^2 + 41 t^2 u^2 +33 t^3 u^2+ 14 t^4 u^2  \cr
&+ 7 u^3 + 21 t u^3 + 33 t^2 u^3 + 30 t^3 u^3+ 14 t^4 u^3  \cr
&+ 7 tu^4 + 14 t^2 u^4 + 14 t^3 u^4 + 7 t^4 u^4\cr
A_6(t,u)&=42 + 105 t + 135t^2 + 105 t^3 + 42 t^4 \cr
&+ 105 u + 275 t u + 375 t^2 u + 322 t^3 u + 168 t^4 u + 42 t^5 u \cr
&+ 135 u^2 + 375 t u^2 + 547 t^2 u^2 + 518 t^3 u^2 + 322 t^4 u^2 + 
105 t^5 u^2 \cr
&+ 105 u^3 +322 t u^3 +518 t^2 u^3 +547 t^3u^3 +375 t^4 u^3 +135 t^5 u^3 \cr
&+ 42 u^4 +168 t u^4 + 322 t^2 u^4 + 375 t^3 u^4 + 275 t^4 u^4 +105 t^5 u^4 \cr
&+ 42 t u^5 + 105 t^2 u^5 + 135t^3u^5 +105 t^4 u^5 + 42 t^5 u^5\cr}}

\vfill\eject

\appendix{B}{Perron-Frobenius vectors for the two-dislocation inhomogeneous O(1) loop model}

We list below the vectors $\Psi_n(t,u)\equiv \Psi_n^{(1,n+1)}(t,u)$, computed
by solving eq.\pfrosimp, up to $n=5$. Like in \DIF, the entries are listed in 
lexicographic order on the link patterns.
\eqn\psinvalues{\eqalign{
&\Psi_1(t,u)=\{1\}\cr
&\Psi_2(t,u)=\{1,tu\}\cr
&\Psi_3(t,u)=\{u(1 + t u), 1, t u, t (1 + t u), t^2 u^2\}\cr
&\Psi_4(t,u)=\{1 + t + u + t u + t^2 u + t u^2 + t^2 u^2, u(1 + t + t u), 
u^2(1 + t u + t^2 u), u (1 + u + t u^2), \cr
&1, t (1 + u + t u), t u, t u^2(1 + t + t u), 
t u (1 + t + u + t u + t^2 u + t u^2 + t^2 u^2), 
t (1 + t + t^2 u), \cr
&t^2u^2, t^2 u(1 + u + t u), 
t^2 (1 + t u + t u^2), t^3 u^3\}\cr
&\Psi_5(t,u)=\{
u(2 + 3 t + 2 t^2 + 3 u + 5 t u + 4 t^2 u + 2 t^3 u + 
2 u^2 + 4 t u^2 + 5 t^2 u^2 + 3 t^3 u^2 + 2 t u^3 + 
3 t^2 u^3\cr
&+ 2 t^3 u^3),
u^2(2 + 3 t + 2 t^2 + u + 3 t u + 3 t^2 u + t u^2 + 
2 t^2 u^2), 2 + 3 t + 2 t^2 + u + 2 t u + 2 t^2 u \cr
&+ 2 t^3 u + t u^2 + 
t^2 u^2 + t^3 u^2,1 + 2 t + 2 u + 4 t u + 2 t^2 u + 2 t u^2 + t^2 u^2, 
u(1 + 2 t + t u), \cr
&u(1 + t + t^2 + 2 u + 2 t u + 2 t^2 u + t^3 u + 2 t u^2 + 
3 t^2 u^2 + 2 t^3 u^2),u^2(1 + t + t^2 + t u + 2 t^2 u), \cr
&1 + t + t^2 + u + t u + t^2 u + t^3 u + u^2 + t u^2 + t^2 u^2 + 
t^3 u^2 + t u^3 + t^2 u^3 + t^3 u^3, \cr
&2 + t + 3 u + 2 t u + t^2 u + 2 u^2 + 2 t u^2 + t^2 u^2 + 
2 t u^3 + t^2 u^3, u(2 + t + u + t u + t u^2), \cr
&u^3(1 + t u + t^2 u + t^3 u), 
u^2(1 + 2 u + 2 t u^2 + t^2 u^2), 
u(1 + u + u^2 + t u^3), 1, \cr
&t u(2 + t + 3 u + 2 t u + t^2 u + 2 u^2 + 2 t u^2 + 
t^2 u^2 + 2 t u^3 + t^2 u^3), 
t u^2(2 + t + u + t u + t u^2), \cr
&t(2 + t + u + t u + t^2 u), t(1 + 2 u + t u), t u, 
t u(1 + 2 t + 2 u + 4 t u + 2 t^2 u + 2 t u^2 + 
t^2 u^2), \cr
&t u^2(1 + 2 t + t u),
t(1 + 2 t + u + 2 t u + 2 t^2 u + u^2 + 2 t u^2 + 
3 t^2 u^2 + t u^3 + 2 t^2 u^3), \cr
&t(2 + 3 t + 2 t^2 + 3 u + 5 t u + 4 t^2 u + 2 t^3 u + 
2 u^2 + 4 t u^2 + 5 t^2 u^2 + 3 t^3 u^2 + 2 t u^3 + 
3 t^2 u^3 + 2 t^3 u^3), \cr
&t u(2 + 3 t + 2 t^2 + u + 2 t u + 2 t^2 u + 2 t^3 u + 
t u^2 + t^2 u^2 + t^3 u^2),t u^3(1 + t + t^2 + t u + 2 t^2 u), \cr
&t u^2(1 + t + t^2 + 2 u + 2 t u + 2 t^2 u + t^3 u + 
2 t u^2 + 3 t^2 u^2 + 2 t^3 u^2),t u(1 + t + t^2 + u + t u + t^2 u \cr
&+ t^3 u + u^2 + t u^2 + 
t^2 u^2 + t^3 u^2 + t u^3 + t^2 u^3 + t^3 u^3),t(1 + t + t^2 + t^3 u), 
t^2 u(1 + 2 u + t u),t^2 u^2, \cr
&t^2(1 + u + t u + u^2 + 2 t u^2),t^2(2 + t + 3 u + 3 t u + t^2 u + 
2 u^2 + 3 t u^2 + 2 t^2 u^2), t^2 u(2 + t + u + t u \cr
&+ t^2 u),t^2 u^3(1 + 2 t + t u), 
t^2 u^2(1 + 2 t + 2 u + 4 t u + 2 t^2 u + 2 t u^2 + t^2 u^2),t^2 u(1 + 
2 t + u + 2 t u  \cr
&+ 2 t^2 u + u^2 + 2 t u^2 + 3 t^2 u^2 + t u^3 + 2 t^2 u^3),t^2(1 + 2 t + 2 t^2 u + 
t^2 u^2), t^3 u^3, t^3 u^2(1 + 2 u + t u), \cr
&t^3 u(1 + u + t u + u^2 + 2 t u^2), 
t^3(1 + t u + t u^2 + t u^3), t^4 u^4\}\cr}}

\appendix{C}{Polynomials for the multiple-dislocation O(1) loop model}
We list below the polynomials $P_n(m\vert t)$ obtained by solving 
eq.\pfrosimplissimus, for $1\leq m\leq n\leq 6$.
\eqn\folpol{\eqalign{
P_1(1\vert t)&=1\cr
P_2(1\vert t)&=1+t\cr
P_2(2\vert t)&=1+t^2\cr
P_3(1\vert t)&=2+3t+2t^2\cr
P_3(2\vert t)&=1+2t+t^2+2t^3+t^4\cr
P_3(3\vert t)&=1+t+3t^2+t^3+t^4\cr
P_4(1\vert t)&=7+14t+14t^2+7t^3\cr
P_4(2\vert t)&=2+6t+9t^2+8t^3+9t^4+6t^5+2t^6\cr
P_4(3\vert t)&=1+4t+6t^2+10t^3+10t^4+6t^5+4t^6+t^7\cr
P_4(4\vert t)&=1+2t+7t^2+6t^3+10t^4+6t^5+7t^6+2t^7+t^8\cr
P_5(1\vert t)&=42 + 105 t + 135 t^2 + 105 t^3 + 42 t^4\cr
P_5(2\vert t)&=7 + 28 t + 58 t^2 + 80 t^3 + 83 t^4 + 80 t^5 + 58 t^6 + 
    28 t^7 + 7 t^8\cr
P_5(3\vert t)&=2 + 11 t + 30 t^2 + 52 t^3 + 76 t^4 + 87 t^5 + 76 t^6 + 
    52 t^7 + 30 t^8 + 11 t^9 + 2 t^{10}\cr
P_5(4\vert t)&=1 + 6 t + 15 t^2 + 34 t^3 + 54 t^4 + 66 t^5 + 77 t^6 + 
    66 t^7 + 54 t^8 + 34 t^9 + 15 t^{10} + 6 t^{11} + t^{12}\cr
P_5(5\vert t)&=1 + 4 t + 16 t^2 + 30 t^3 + 55 t^4 + 67 t^5 + 83 t^6 + 
    67 t^7 + 55 t^8 + 30 t^9 + 16 t^{10} + 4 t^{11} + t^{12}\cr
P_6(1\vert t)&=429 + 1287 t + 2002 t^2 + 2002 t^3 + 1287 t^4 + 429 t^5 \cr 
P_6(2\vert t)&=42 + 210 t + 545 t^2 + 960 t^3 + 1275 t^4 + 1372 t^5 + 
1275 t^6 + 960 t^7 + 545 t^8 + 210 t^9 + 42 t^{10} \cr
P_6(3\vert t)&= 7 + 49 t + 174 t^2 + 412 t^3 + 730 t^4 + 1062 t^5 + 1284 t^6 + 
1284 t^7 + 1062 t^8 + 730 t^9 + 412 t^{10} \cr
&+ 174 t^{11} + 49 t^{12} + 7 t^{13}\cr
P_6(4\vert t)&= 2 + 16 t + 64 t^2 + 168 t^3 + 354 t^4 + 608 t^5 + 868 t^6 + 
1064 t^7 + 1148 t^8 + 1064 t^9 + 868 t^{10} \cr
&+ 608 t^{11} + 
354 t^{12} + 168 t^{13} + 64 t^{14} + 16 t^{15} + 2 t^{16}\cr
P_6(5\vert t)&= 1 + 9 t + 36 t^2 + 110 t^3 + 255 t^4 + 467 t^5 + 738 t^6 + 
972 t^7 + 1130 t^8 + 1130 t^9 + 972 t^{10} \cr
&+ 738 t^{11} + 
467 t^{12} + 255 t^{13} + 110 t^{14} + 36 t^{15} + 9 t^{16} + t^{17}\cr
P_6(6\vert t)&= 1 + 6 t + 30 t^2 + 84 t^3 + 204 t^4 + 372 t^5 + 624 t^6 + 
828 t^7 + 1035 t^8 + 1068 t^9 + 1035 t^{10} \cr
&+ 828 t^{11} + 
624 t^{12} + 372 t^{13} + 204 t^{14} + 84 t^{15} + 30 t^{16} + 6 t^{17} + 
t^{18}\cr}}

\appendix{D}{Polynomials for the O(1) loop model on a strip of finite width}

Up to $m=10$, the polynomials $P_m^O(t)$ \sumeven-\sumodd\ read:
\eqn\oppol{\eqalign{
P_{1}^O(t)&=1\cr
P_{2}^O(t)&=1\cr
P_{3}^O(t)&=1+t\cr
P_{4}^O(t)&=1+t+t^2\cr
P_{5}^O(t)&=1 + 3 t + 3 t^2 + 3 t^3 + t^4\cr
P_{6}^O(t)&=1+3t+6t^2+6t^3+6t^4+3t^5+t^6\cr
P_{7}^O(t)&=1+6t+15t^2+28t^3+35t^4+35t^5+28t^6+15t^7+6t^8+t^9\cr
P_{8}^O(t)&=1 + 6 t + 21 t^2 + 46 t^3 + 81 t^4 + 108 t^5 + 120 t^6 + 
108 t^7+81 t^8 + 46 t^9 + 21 t^{10} + 6 t^{11} + t^{12}\cr
P_9^O(t)&=1 + 10 t + 45 t^2 + 140 t^3 + 320 t^4 + 585 t^5 + 886 t^6 + 
1120 t^7 + 1215 t^8 + 1120 t^9 + 886 t^{10} \cr
&+585 t^{11}+320 t^{12} + 140 t^{13} + 45 t^{14} + 10 t^{15} + t^{16}\cr
P_{10}^O(t)&=1 + 10 t + 55 t^2 + 200 t^3 + 560 t^4 + 1253 t^5 + 
2345 t^6+3740 t^7 + 5180 t^8 + 6260 t^9 + 6677 t^{10}\cr
&+ 6260 t^{11}+5180 t^{12}+3740 t^{13} + 2345 t^{14} + 1253 t^{15} + 560 t^{16} + 200 t^{17} + 
55 t^{18}+10 t^{19} + t^{20}\cr}}

\listrefs

\bye